\documentclass[aps, pra, 10pt, longbibliography]{revtex4-2}
\usepackage{graphicx}
\usepackage{amsmath}
\usepackage{xcolor}
\usepackage{soul}
\usepackage{multirow}
\definecolor{OliveGreen}{RGB}{0,200,0}

\newcommand{\re}{Re_{\tau}}
\newcommand{\rev}[1]{{\color{black}#1}}
\newcommand{\rahul}[1]{{\color{black}#1}}

\begin{document}

\title{A note on the amplitude modulation phenomenon\\in non-canonical wall-bounded flows}

\author{Mitchell Lozier}
\email[]{mlozi2018@gmail.com}
\author{Ivan Marusic}
\author{Rahul Deshpande}
\affiliation{Department of Mechanical Engineering, The University of Melbourne, Victoria 3010, Australia}

\date{\today}

\begin{abstract}
The amplitude modulation phenomena, defined originally by Mathis \emph{et al.} ({J.\ Fluid Mech.}, \textbf{628}, 311-337; 2009), corresponds to a unique non-linear interaction between Reynolds number ($\re$) dependent large-scale motions and $\re$-invariant inner-scale motions observed in canonical wall-bounded flows. 
While similar non-linear interactions \rev{have been quantified previously} in non-canonical wall-bounded flows, linking them \rev{solely} to amplitude modulation is questionable due to the fact that each non-canonical effect is associated with distinct variations in the energies of both the large and inner scaled motions. 
This study \rev{revisits analysis of} non-linear triadic interactions, \rev{with consideration to various non-canonical effects,} by analyzing published hot-wire datasets acquired in the large Melbourne wind tunnel. 
It is found that triadic interactions, across the entire turbulence scale hierarchy, \rev{may} become statistically significant with increasing intensity of non-canonical effects such as wall roughness, pressure gradients, and spanwise or wall-normal forcing \rahul{(when compared relative to their respective canonical baseline cases at matched $\re$)}. 
This stands in contrast to previous observations made in canonical flows, where only the interaction between inner scales and inertia-dominated large scales was considered dynamically significant for increasing $\re$. 
The implications of these findings are discussed for near-wall flow prediction models in non-canonical flows, which \rev{should take into account \emph{all} non-linear interactions coexisting in wall-bounded flows}. 
\end{abstract}

\maketitle


\section{Introduction and motivation}

A characteristic feature of all turbulent flows is a broad range of turbulent scales of motion, which are non-linearly coupled across the energy spectrum. 
This inter-scale coupling/interaction regulates the energy transfer mechanisms that drive several technologically relevant flows, such as the turbulent boundary layer (TBL). 
TBLs encountered in engineering applications are typically exposed to non-canonical effects/perturbations, such as wall roughness, pressure gradients, etc., which affect these inter-scale interactions in a non-trivial manner. 
Here we will discuss a well-known inter-scale interaction, \emph{i.e.} the amplitude modulation phenomenon \citep{bandyopadhyay1984,mathis_large-scale_2009}, in the context of non-canonical wall-bounded flows. 
Before investigating non-canonical effects, however, we will first define the various terminologies and revisit the established knowledge on inter-scale interactions in a canonical TBL, \emph{i.e.} a zero-pressure gradient (ZPG) TBL over a hydraulically-smooth wall. 

\begin{figure}
\includegraphics[width=0.6\textwidth]{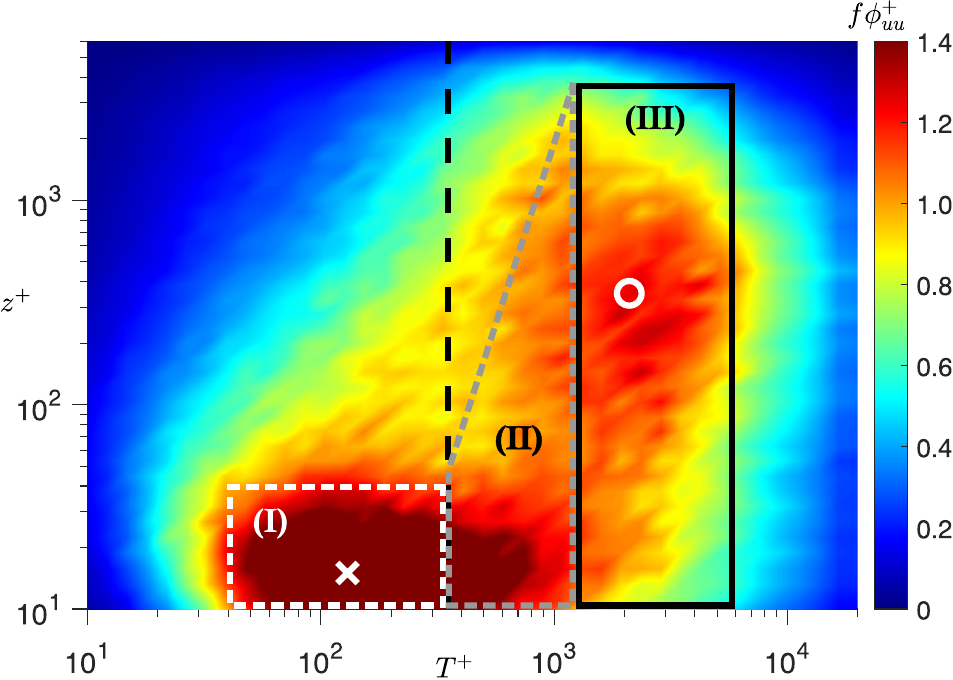}
\caption{\label{Fig0} 
Premultiplied energy spectra of a canonical TBL at $\re=8100$ from Marusic et al. \cite{marusic_evolution_2015}. Vertical black dashed line represents the cut-off timescale $T^{+}_{c}=350$. The white $\times$ and $\circ$ mark the approximate `inner' and `outer' peaks respectively.}
\end{figure}

Figure~\ref{Fig0} depicts a spectral representation of the energy distribution across a broad hierarchy of scales coexisting in a high-Reynolds number ($\re$ $>$ $\mathcal{O}$($10^3$)) canonical TBL \cite{marusic_evolution_2015}. 
It is presented as the premultiplied frequency ($f$) spectra of the streamwise velocity fluctuations ($f\phi^{+}_{uu}$) plotted for various wall-normal distances ($z^{+}=zU_{\tau}/\nu$) and as a function of timescales ($T^{+}=TU^{2}_{\tau}/\nu$ = $U^{2}_{\tau}/{f{\nu}}$). 
Here, $U_{\tau}$ and $\nu$ denote the mean friction velocity and kinematic viscosity used to normalize flow properties in viscous scaling (indicated with the superscript `$+$'). 
The Reynolds number of the TBL will be quantified using the friction Reynolds number, $\re=U_{\tau}\delta/\nu$, where $\delta$ is the TBL thickness. 
In figure~\ref{Fig0}, the outlined region (I) captures the energy distribution across scales associated with the $\re$-invariant near-wall cycle \cite{kline_structure_1967, hutchins_evidence_2007}, and the associated `inner' peak location is marked by a white $\times$ for reference. 
Slightly larger timescales, outlined in region (II), correspond to a group of eddies/motions exhibiting distance-from-the-wall scaling (e.g. uniform momentum zones \cite{adrian_vortex_2000} or attached/wall-scaled eddies \cite{hwang_wall-attached_2018, hu_wall-attached_2020}). 
The overall energy associated with region (II) is Reynolds number dependent, owing to the growth in the hierarchy/range of wall-scaled eddies with increasing $\re$ \cite{hwang_wall-attached_2018}.
The final outlined region (III) is associated with large $\delta$-scaled eddies or superstructures, whose energies are also dependent on the Reynolds number \cite{hutchins_evidence_2007,lee_very-large-scale_2011, dennis_experimental_2011}. 
This region exhibits a peak at sufficiently high $Re_{\tau}$, the approximate location for which has been marked by a white $\circ$ for reference. 
A cut-off timescale of $T_{C}^{+}=350$ (vertical dashed line in figure~\ref{Fig0}) has often been used in the literature to nominally separate these Reynolds number dependent ($u_{L}$) and invariant ($u_{i}$) motions \cite{deshpande_relationship_2023}, a distinction which will be used throughout the present study to decompose $u$-fluctuations into $u_{L} = u(T^{+}$ $\gtrsim$ $T^{+}_{c})$ and $u_{i} = u(T^{+}<T^{+}_{C})$. 
Previous studies have generally used the terminology pairings of large- and small-scale \emph{or} inner- and outer-scale to differentiate between these decomposed fluctuations (see \citep{mathis_large-scale_2009} and \citep{deshpande_relationship_2023}, respectively, for example). 
In the present study we will use the terminology `large-scale' to refer to $u_{L} = u(T^{+}$ $\gtrsim$ $T^{+}_{c})$ and `inner-scale' to refer to $u_{i} = u(T^{+}<T^{+}_{C})$.

For a canonical TBL, it is now well-established that the variance of the velocity fluctuations associated with the inner scales ($\overline{u_{i}^{+^{2}}}$, corresponding to region (I) of figure~\ref{Fig0}), does not vary with $\re$ \citep{hutchins_hot-wire_2009}. 
Conversely, the variance of large-scale velocity fluctuations ($\overline{u_{L}^{+^{2}}}$ corresponding to regions (II) and (III) of figure~\ref{Fig0}), increases with $\re$ throughout the TBL. 
Hence, $\overline{u_{L}^{+^{2}}}$ is solely responsible for the observed $\re$-increase in near-wall streamwise turbulence intensity in a canonical TBL \cite{hutchins_hot-wire_2009}. 
Hutchins \& Marusic \cite{hutchins_evidence_2007} described these increasingly energetic large-scale signatures, in the near-wall region, to be a superimposition (akin to a `footprint') of the structures originating in the outer region of the TBL. 
They also observed a large-scale modulating effect on the amplitude of small-scale (inner-scale) velocity fluctuations in the near-wall region, which was found to be reminiscent of pure amplitude modulation, and was referred to as such. 
Mathis et al. \cite{mathis_large-scale_2009} introduced an amplitude modulation coefficient ($R$) to quantify this modulation of the inner scales, by the large scales, at any wall-normal location $z$. 
$R$ was defined as: 
\begin{equation}
R(z) = \frac{\overline{u_{L}^{+}(z)E_{L}(u_{i}^{+}(z))}}{{\sqrt{\overline{u_{L}^{2+}(z)}}\ \sqrt{\overline{E_{L}(u_{i}^{+}(z))^2}}}},
\label{eq:R}
\end{equation}
\noindent where $u_{L}^{+}$ represents large-scale streamwise velocity fluctuations and $E_{L}(u_{i}^{+})$ represents a large-scale filtered envelope of $u_{i}$ computed via the Hilbert transform. 
Although it is agreed that two-point correlation methods give an overall higher fidelity quantification of amplitude modulation \cite{bernardini_innerouter_2011, jacobi_phase_2013, dogan_quantification_2019}, Mathis et al. \citep{mathis_large-scale_2009} demonstrated the single-point correlation coefficient in Eq. (\ref{eq:R}) yields a qualitatively similar result to the two-point correlation under most conditions.  

Here, it is important to note the fact that the amplitude modulation phenomena was originally proposed in the context of a canonical high-$\re$ TBL, to describe exclusively the effect of energetic large-scale structures superimposing onto, and modulating the amplitude of, inner scales (which are $\re$-invariant). 
Per this original definition, the phenomenon is associated with an inherent sense of directionality, \emph{i.e.} it is implied that the large-scale structure dynamics govern their non-linear interaction with the inner scales. 
This makes sense given that the experiments of Hutchins \& Marusic \cite{hutchins_evidence_2007} and Mathis et al. \citep{mathis_large-scale_2009} were focused on $\re$-effects in a canonical TBL, where the increase in amplitude modulation can be solely attributed to the $\re$-dependent energization of the large-scale structures. 
It is also noteworthy that the existence/description of causality between large-scale superimposition and amplitude modulation is still a topic of ongoing research \cite{andreolli_separating_2023}. 
Previous studies have also \rahul{proposed} alternative descriptions of the amplitude modulation phenomena for canonical TBLs, such as the quasi-steady quasi-homogeneous (QSQH) theory \rahul{of} Chernyshenko and co-workers \citep{zhang2016qsqh, chernyshenko2021extension}. 
This theory mathematically relates the amplitude of inner-scale fluctuations in the near-wall region to large-scale fluctuations in skin friction, without making explicit assumptions about the nature of interactions between structures in the inner and outer layers (\emph{i.e.} superimposition). 
As such, the `cause-and-effect' relationship between amplitude modulation and large-scale motions in the near-wall region remains an open question. 
\rahul{It is noteworthy to also consider the recent interest in revisiting the physical interpretation of the amplitude modulation coefficient definition ($R$), based on consideration of the interactions between the mean flow and the large-scales \cite{Cui_AmplitudeModulation_}. 
However, the present study limits its investigation and interpretation to the classical definitions of quantifying the hierarchy of non-linear triadic interactions within the TBL (amplitude modulation included; \cite{mathis_large-scale_2009,duvvuri_triadic_2015}), and extending it to the TBL scenarios associated with different non-canonical perturbations. 
As a consequence, the present conclusions do not depend on any new perceptions/interpretations of the exact physical mechanisms governing the amplitude modulation phenomena.} 

Based on the observations described above, the amplitude modulation phenomena can be considered as a subset of the non-linear triadic interactions coexisting across the entire TBL scale hierarchy \cite{duvvuri_triadic_2015}. 
The triadic nature of these interactions emerges from the quadratic non-linearity in the governing Navier-Stokes equations. 
Triadically-coupled scales correspond to the frequency scales of three eddies (say, $l$, $m$ and $n$) that are inter-connected via any of the following relations: 
\begin{equation}
(a)\ \omega_{l} = \omega_{m} - \omega_{n}, \quad (b)\ \omega_{l} = \omega_{m} + \omega_{n}, \quad (c)\ \omega_{l} = 2 \omega_{m} \quad \textrm{or} \quad (d)\  \omega_{l} = 2 \omega_{n},
\label{eq:triad}
\end{equation}
\noindent where, $\omega_{i}={2\pi{f_{i}}}=2\pi/T_{i}$. 
Per the framework proposed by McKeon and co-workers \cite{jacobi_phase_2013, duvvuri_triadic_2015}, the mean non-linear coupling between all triadically coupled scales (at any $z$) can be quantified by the skewness of the $u$-fluctuations, $S_{u}$. 
This was demonstrated by decomposing statistically stationary streamwise velocity time series, $u(t)$, into Fourier modes: $u(t) = \sum^{\infty}_{i=1} \alpha_{i}sin(\omega_{i}t + \psi_{i})$, with amplitudes $(\alpha_{i})$, phase $(\psi_{i})$ and $0<\omega_{i}<\omega_{\infty}$. 
Hence, 
\begin{equation}
S_{u} = \frac{\overline{u^{3}}}{\sigma^{3}} = \frac{6}{4\sigma^{3}} \sum_{\substack{\forall\ l,m,n \\ \omega_{l}<\omega_{m}<\omega_{n} \\ \omega_{l}+\omega_{m}=\omega_{n}}} \alpha_{l}\alpha_{m}\alpha_{n} sin(\psi_{l}+\psi_{m}-\psi_{n}) + \frac{3}{4\sigma^{3}} \sum_{\substack{l=1\\ \omega_{n}=2\omega_{l}}} \alpha_{l}^{2}\alpha_{n} sin(2\psi_{l}-\psi_{n}),
\label{eq:skew1}
\end{equation}
\noindent where $\sigma=\sqrt{\overline{u^{2}}}$, and $\psi_{l}+\psi_{m}-\psi_{n}$ represents the phase difference between triadically coupled scales. 
Based on Eq. (\ref{eq:skew1}), $S_{u}$ can be considered a surrogate for the average measure of phase difference between triadically coupled scales across the full spectrum of energetic turbulent eddies/motions \citep{duvvuri_triadic_2015}. 
Thus, higher $S_{u}$ can be interpreted as a stronger coupling/interaction owing to a lower phase lag between triadically coupled scales. 
\rev{Consideration of skewness to quantify inter-scale coupling is practical here because it does not require any normalization (by characteristic length or velocity scales), which could otherwise undermine comparisons between datasets covering a broad range of non-canonical characteristics. 
An additional advantage is that the skewness does not require decomposition of the velocity for computation. 
However, inter-scale interactions can be described in more detail by computing the individual (small) inner- ($u_{i} = u(T^{+}<T^{+}_{C})$) and (outer) large- ($u_{L} = u(T^{+}>T^{+}_{c})$) scale contributions to $S_{u}$, where the cut-off timescale considered is $T^{+}_{c} = 2\pi/\omega^{+}_{c} = 350$ \citep{deshpande_relationship_2023}.} 
This decomposition based on frequency cut-off (instead of a wavelength-based cut-off) is preferred, given it avoids dependence on the choice of convective velocity, which may influence the amplitude modulation coefficient \citep{yang_implication_2018}. 
This decomposition enables estimation of the non-linear coupling between $u_{L}$ and $u_{i}$, for which we substitute $u=u_{L}+u_{i}$ into (\ref{eq:skew1}) following Mathis et al. \cite{mathis_relationship_2011} to obtain: 
\begin{equation}
S_{u}=\frac{\overline{u^{3}}}{\sigma^{3}} = \overline{\overline{u^{3}}} =\overline{\overline{u_{L}^{3}}} + \overline{\overline{u_{i}^{3}}} + 3\overline{\overline{u_{L}^{2}u_{i}}} + 3\overline{\overline{u_{i}^{2}u_{L}}} \; ,
\label{eq:skew2}
\end{equation}
\noindent where the double overbar denotes a time-averaged quantity normalized by $\sigma^{3}$. 
Duvvuri and McKeon \cite{duvvuri_triadic_2015} reported exact expressions for the individual terms in Eq. (\ref{eq:skew2}), with 
\begin{equation}
\overline{\overline{u_{i}^{2}u_{L}}} = \frac{1}{2\sigma^{3}} \sum_{\substack{\forall\ l,m,n \\ \omega_{n}-\omega_{m}=\omega_{l} \\ 0<\omega_{l}<\omega_{c} \\ \omega_{m},\omega_{n}>\omega_{c}}} \alpha_{l}\alpha_{m}\alpha_{n} sin(\psi_{l}+\psi_{m}-\psi_{n}) = 
{\frac{R}{2}}{\bigg(}{{\sqrt{\overline{u_{L}^{2+}(z)}}\ \sqrt{\overline{E_{L}(u_{i}^{+}(z))^2}}}}{\bigg)},
\label{eq:interscale}
\end{equation}
\noindent indicating that the `cross-term' $3\overline{\overline{u^{2}_{i}u_{L}}}$ represents the mean phase difference between the large scales $\omega_{l}$ and an `envelope' of the triadically-coupled inner scales: $\omega_{n}$,$\omega_{m}$ ($>$ $\omega_c$). 
Increasing values for this cross-term are therefore associated with a stronger inter-scale coupling/interaction owing to a reduction in phase between $u_{L}$ and $u_{i}$. 
Mathis et al. \citep{mathis_relationship_2011} observed experimentally that $3\overline{\overline{u^{2}_{i}u_{L}}}$ behaved similarly to the correlation coefficient ($R$) used to quantify the amplitude modulation phenomena, and was also the most dominant scale-decomposed skewness term (Eq. \ref{eq:skew2}) for a canonical TBL. 
Additionally, this cross-term had a strong Reynolds number dependence that extended across the whole TBL, akin to the behavior of $R$. 
The directly proportionality between the cross-term and the amplitude modulation coefficient ($R$) was formally established by Duvvuri \& McKeon \cite{duvvuri_triadic_2015} through Eq. (\ref{eq:interscale}), thereby also confirming $R$ to represent a subset of the non-linear interactions coexisting across the full TBL scale hierarchy. 

A number of studies have also used the correlation coefficient from Eq. (\ref{eq:R}) to quantify the inter-scale coupling in non-canonical TBLs. 
The results from these non-canonical studies are often interpreted such that any increment in $R$ is an artifact of amplitude modulation, in the classical sense defined by Mathis et al. \cite{mathis_large-scale_2009} (\emph{i.e.} driven by changes in the large-scale dynamics). 
However, inter-scale coupling measured in most, if not all, non-canonical TBLs can pertain to interactions beyond the specific subset of interactions which correspond to amplitude modulation (defined in Eqs. \ref{eq:R}$-$\ref{eq:interscale}). 
For instance, the classical definition of amplitude modulation would preclude describing changes in inter-scale coupling in flows without an energetic large-scale structure, as amplitude modulation (\emph{i.e.} ZPG TBLs at $Re_{\tau}$ $\lesssim$ $\mathcal{O}$($10^2$) \cite{hutchins_evidence_2007}, where regions (II) and (III) of figure~\ref{Fig0} are statistically insignificant). 
Similarly, it would also be incorrect to describe changes in inter-scale coupling as amplitude modulation if they are solely associated with manipulation of the near-wall cycle (\emph{i.e.} region (I) in figure~\ref{Fig0}), which occurs owing to non-canonical effects such as roughness \cite{anderson_amplitude_2016} or porous surfaces \cite{kim_experimental_2020}. 
Interestingly, other non-canonical effects such as pressure gradients \cite{harun_pressure_2013,deshpande_reynolds-number_2023}, thermal stratification \cite{salesky_buoyancy_2018} or large-scale free-stream disturbances \cite{dogan_interactions_2016} have also been found to be responsible for affecting the entire hierarchy of turbulent scales (\emph{i.e.} across regions (I)-(III) in figure~\ref{Fig0}). 
However, associating the corresponding changes in inter-scale coupling solely with the amplitude modulation phenomena would not be appropriate, given they are likely to be an artifact of changes in both $u_{L}$ and $u_{i}$, differing from the classical definition. 
Additionally, in all the above non-canonical scenarios, the inherent uni-directionality associated with the amplitude modulation phenomena (regarding large-scale structure dynamics affecting the inner scales) may not be strictly applicable. 

\rev{These observations suggest that interpretation of the correlations conventionally associated with amplitude modulation (\emph{i.e.} $R$ and $3\overline{\overline{u^{2}_{i}u_{L}}}$) should be considered \emph{cautiously}, especially in the context of non-canonical wall-bounded flows. 
To emphasize this, the current study will focus on quantifying changes across the hierarchy of inter-scale interactions, which arise from a wide variety of non-canonical perturbations, \rahul{by demonstrating the non-trivial effect of these perturbations relative to their canonical baseline cases at matched $Re_{\tau}$}. 
Specific instances will be identified empirically where $3\overline{\overline{u^{2}_{i}u_{L}}}$ (or $R$) alone are insufficient to draw conclusions regarding changes in amplitude modulation phenomena, owing to some non-canonical effects as described qualitatively above. 
The implications of these findings with respect to the modeling of wall-bounded flows will also be discussed throughout. 
\rahul{In that sense, the present findings motivate further investigations towards the appropriate quantification/description of the} amplitude modulation phenomena in the future, with particular emphasis on improving analysis/modeling of non-canonical wall-bounded flows.} \\

\section{Experimental datasets}

\rev{In the present study, we argue that changes in the modulation coefficient, $R$, or the cross-term, $3\overline{\overline{u^{2}_{i}u_{L}}}$, alone are insufficient to draw conclusions regarding amplitude modulation phenomena in the context of non-canonical TBLs. 
For this, we have assembled and analyzed a unique set of published hot-wire datasets from a single experimental facility, the large Melbourne wind tunnel. 
These datasets will be used independently to demonstrate enhancement of a broad range of inter-scale couplings} \rahul{(owing to their non-canonical perturbations), by comparing with their respective canonical baseline cases at matched $Re_{\tau}$}.
The measurement details of all these datasets can be found in their respective references, which have been documented alongside the range of their respective perturbations in table \ref{Tab1}. 
Each of these datasets comprise a canonical baseline case, \emph{i.e.} corresponding to a high-$\re$ ZPG TBL over a smooth wall, against which the effect of each perturbation, on inter-scale coupling, will be analyzed at matched TBL $Re_{\tau}$. 
A unique aspect of these datasets is that each of them is acquired with nominally similar hot-wire resolution $l^+$, ensuring that the effect of spatial filtering emerging from finite hot-wire length ($l$; \cite{hutchins_hot-wire_2009}) will not adversely affect our conclusions. 
To that end, the datasets in table~\ref{Tab1} have been re-analyzed using the framework described in Eqs. (\ref{eq:R})$-$(\ref{eq:interscale}) and the compiled results are presented in figures~\ref{Fig2a} and \ref{Fig3}. 
The normalized inner- and large-scale contributions to the variance of streamwise velocity fluctuations ($\overline{u_{i}^{2^{+}}}$ and $\overline{u_{L}^{2^{+}}}$, respectively), estimated based on $T^+_C$ = 350, will be used to identify the scale range and wall-normal regions that are affected by each perturbation. 
These conclusions, however, should not be influenced by the choice of a viscous-scaled frequency cut-off ($T^+_C$; \citep{mathis_large-scale_2009}) since every non-canonical case is compared against its canonical baseline case at matched $\re$ (refer to table \ref{Tab1}). 
\rev{By applying a consistent cut-off for all cases, observations regarding how each non-canonical perturbation changes the distribution of energy, relative to the canonical baseline case (where figure~\ref{Fig0} is representative of all canonical baseline cases), remains unbiased and consistent with analysis reported in the literature.} 
Complimenting the variance, the premultiplied energy spectra from the near-wall and outer regions will also be presented to corroborate our discussion based on $\overline{u_{i}^{2^{+}}}$ and $\overline{u_{L}^{2^{+}}}$. 
All these statistics, as well as their reference $z$-locations, have been normalized using the friction velocities ($U_{\tau}$) associated with the respective perturbed cases. 
\rev{The choice of $U_{\tau}$, however, does not significantly affect the main conclusions that will be based predominantly on the skewness terms Eq. (\ref{eq:skew2}). 
These terms are considered to quantify the effect of non-canonical perturbations on the various triadic interactions coexisting in the TBL.} 
Particular attention is given to the near-wall region since several past efforts have focused on leveraging inter-scale interactions to predict the turbulence characteristics in this region \citep{marusic_predictive_2010}, which is often inaccessible in high-$\re$ experiments. 
In figures \ref{Fig2a} and \ref{Fig3}, variances and non-linear coupling terms associated with the canonical baseline cases have been plotted using light shading and a solid black outline, for all cases considered in table \ref{Tab1}. \\

\begin{table}
\caption{\label{Tab1} Published hot-wire datasets considered in the present study}
\begin{ruledtabular}
\begin{tabular}{ccccc}
Hot-Wire Dataset & Baseline Canonical TBL & Perturbation & Magnitude & Hot-wire resolution \\
\hline
Marusic et al. \cite{marusic_evolution_2015} & Smooth-Wall ZPG, $Re_{\tau}\approx2800$ & Increasing $Re_{\tau}$ & $2800<Re_{\tau}<13400$ & $l^{+}\sim24$ \\
Deshpande et al. \cite{deshpande_reynolds-number_2023} & Smooth-Wall ZPG, $Re_{\tau}\approx6500$ & Increasing APG & $0.0<\beta<1.7$ & $l^{+}\sim10$ \\
Squire et al. \cite{squire_comparison_2016} & Smooth-Wall ZPG, $Re_{\tau}\approx6500$ & Increasing surface & $0<k_{s}^{+}<97$ & $l^{+}\sim22$ \\
 & & roughness & & \\
Deshpande et al. \cite{deshpande_relationship_2023} & Smooth-Wall ZPG, $Re_{\tau}\approx6000$ & Increasing spanwise & 0.0 ${\le}\;{A^{+}}\;{\le}$ 12.3\:, & $l^{+}\sim8$ \\
 & & wall forcing & 0.000 ${\le}\;{{f}^{+}_{osc}}\;{\le}$ 0.007 & \\
Abbassi et al. \cite{abbassi_skin-friction_2017} & Smooth-Wall ZPG, $Re_{\tau}\approx14400$ & $u_{L}$ Opposing jet & $-$ & $l^{+}\sim20$ \\
Abbassi et al. \cite{abbassi_skin-friction_2017} & Smooth-Wall ZPG, $Re_{\tau}\approx14400$ & $u_{L}$ Reinforcing jet & $-$ & $l^{+}\sim20$ \\
\end{tabular}
\end{ruledtabular}
\end{table}

\section{Non-linear interactions for varying perturbations}

\subsection{Increasing Reynolds number, $Re_{\tau}$}

The first dataset is from the canonical TBL study of Marusic et al. \cite{marusic_evolution_2015}, where the perturbation of interest is an increase in the flow $\re$. 
In this experiment, increasing $\re$ values were achieved by maintaining a nominally constant upstream flow condition and conducting hot-wire measurements at various downstream locations, \rev{resulting in} an increase in $\re$ without significant changes to the viscous scale. 
It is noteworthy that this specific perturbation informed the original description of amplitude modulation by Mathis et al. \cite{mathis_large-scale_2009}, where the $\re$-increase in the energy of large-scale velocity fluctuations in regions (II) and (III) (of figure~\ref{Fig0}) led to an increase in the coupling (reduction in phase difference) between the inner- and large-scaled velocity fluctuations \cite{mathis_large-scale_2009,deshpande_relationship_2023}. 

\begin{figure}
\includegraphics[width=0.81\textwidth]{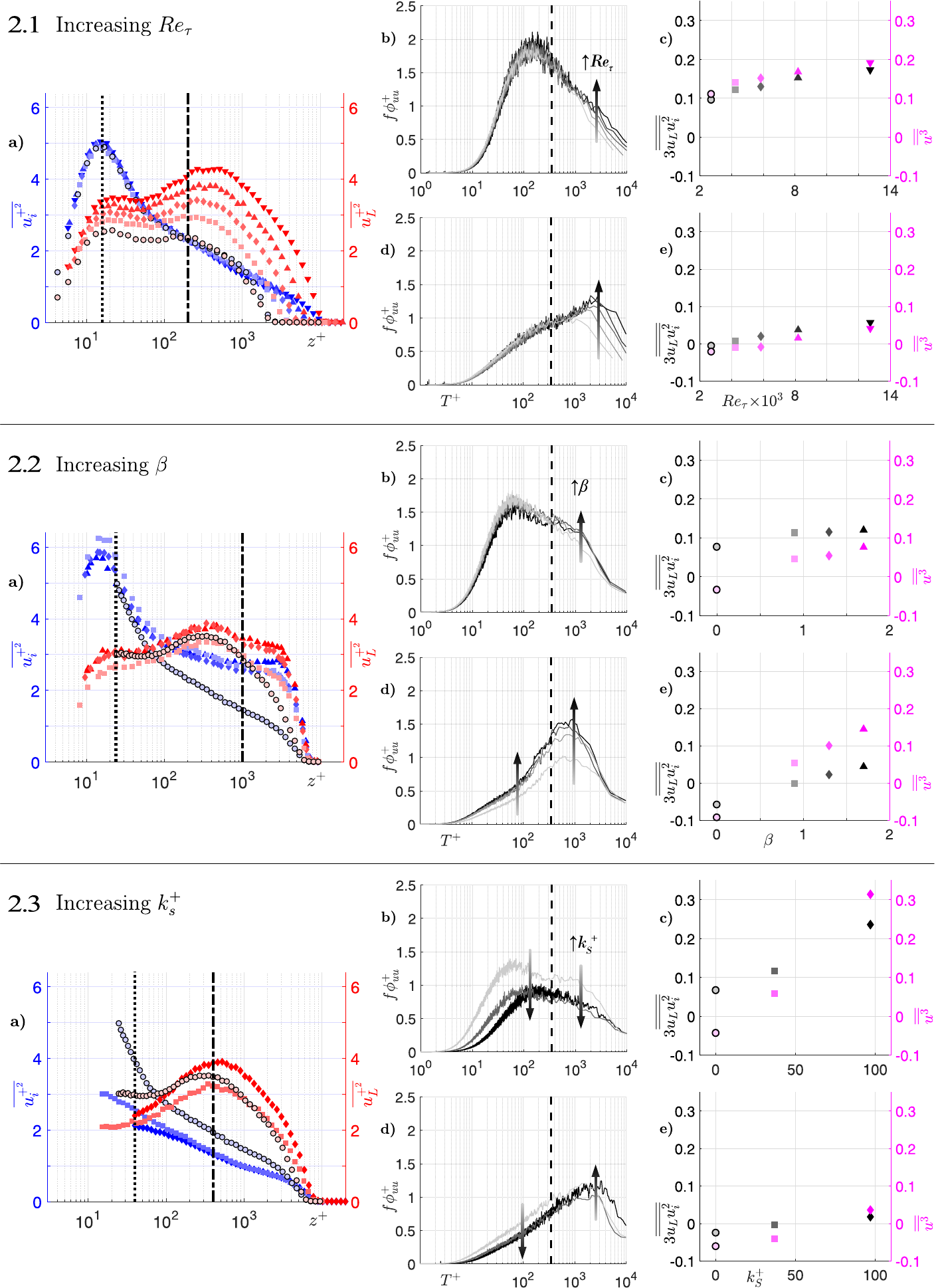}
\caption{\label{Fig2a}
(\emph{left column}) Variance of inner- (left axis) and large- (right axis) scale velocity fluctuations for varying intensities of the following perturbations: \textbf{2.1} Increasing $Re_{\tau}$, \textbf{2.2} Increasing $\beta$ and \textbf{2.3} Increasing $k^+_s$. 
(\emph{center column}) Premultiplied energy spectra for $z^+$-locations corresponding to (b) the inner region (indicated by a vertical dotted line in (a)) and (d) the outer region (indicated by a dash-dotted line in (a)). 
Vertical dashed lines in the center column represent the cut-off timescale, $T^{+}_{c}=350$. 
(\emph{right column}) Cross-term, $3\overline{\overline{u_{L} u_{i}^{2}}}$, (left axis) and skewness of streamwise velocity fluctuations, $\overline{\overline{u^{3}}}$ (right axis) as a function of increasing perturbation intensities at $z^+$-locations, matched with (b,d) respectively, corresponding to (c) the inner region and (e) the outer region. 
Reference canonical cases are shown in the lightest shading, and symbols for the reference canonical cases have a solid black outline.}
\end{figure}

Statistics from the Marusic et al. \cite{marusic_evolution_2015} dataset are presented in figures~\ref{Fig2a}.1. 
As expected based on the literature \citep{hutchins_hot-wire_2009}, there is no change in $\overline{u_{i}^{+^{2}}}$ with increasing $\re$ (light to dark colors in figure~\ref{Fig2a}.1a), while a significant change in $\overline{u_{L}^{+^{2}}}$ can be seen across all $z^{+}$-locations. 
\rev{This trend also gives support to the choice of $T_C^+=350$, as the invariant behavior of $\overline{u_{i}^{+^{2}}}$ with increasing $\re$ has been captured, consistent with expectations \citep{hutchins_hot-wire_2009}.} 
The premultiplied spectra in figure~\ref{Fig2a}.1b are from $z^{+}=15$, as indicated by the vertical dotted line in figure~\ref{Fig2a}.1a. 
The only significant change in the spectra is an increase in energy corresponding to the large scales (right of vertical dotted line at $T^{+}_{C}=350$), \emph{i.e.} corresponding to an increase in $u_{L}$-energy. 
A similar effect is seen in the premultiplied spectra in figure~\ref{Fig2a}.1d which are from $z^{+}=200$ (\emph{i.e.} \rev{nominally within the log region}), indicated by the vertical dash-dotted line in figure~\ref{Fig2a}.1a. 
The effect of increasing $\re$ on the skewness and the cross-term can be seen in figures~\ref{Fig2a}.1c,e. 
Consistent positive changes in the skewness and cross-term, with increasing $\re$, are observed for both $z$-locations. 
\rev{Though not shown here, this same trend can be observed in the skewness at wall-normal locations across the log region \citep{mathis_relationship_2011}.}
The assertion that amplitude modulation (\emph{i.e.} the cross-term) is the only statistically significant triadic/non-linear interaction for a canonical TBL is confirmed by the similar magnitudes of the skewness and the cross-term in figures~\ref{Fig2a}.1c,e. 
This is reaffirmed in figures~\ref{Fig3}.1a,b which demonstrate a negligible variation in all other terms of Eq. (\ref{eq:skew2}) with increasing $\re$ (also shown by \citep{mathis_relationship_2011}), thereby suggesting insignificant changes in other triadic interactions with increasing $\re$. \\

\addtocounter{figure}{-1}
\begin{figure}
\includegraphics[width=0.81\textwidth]{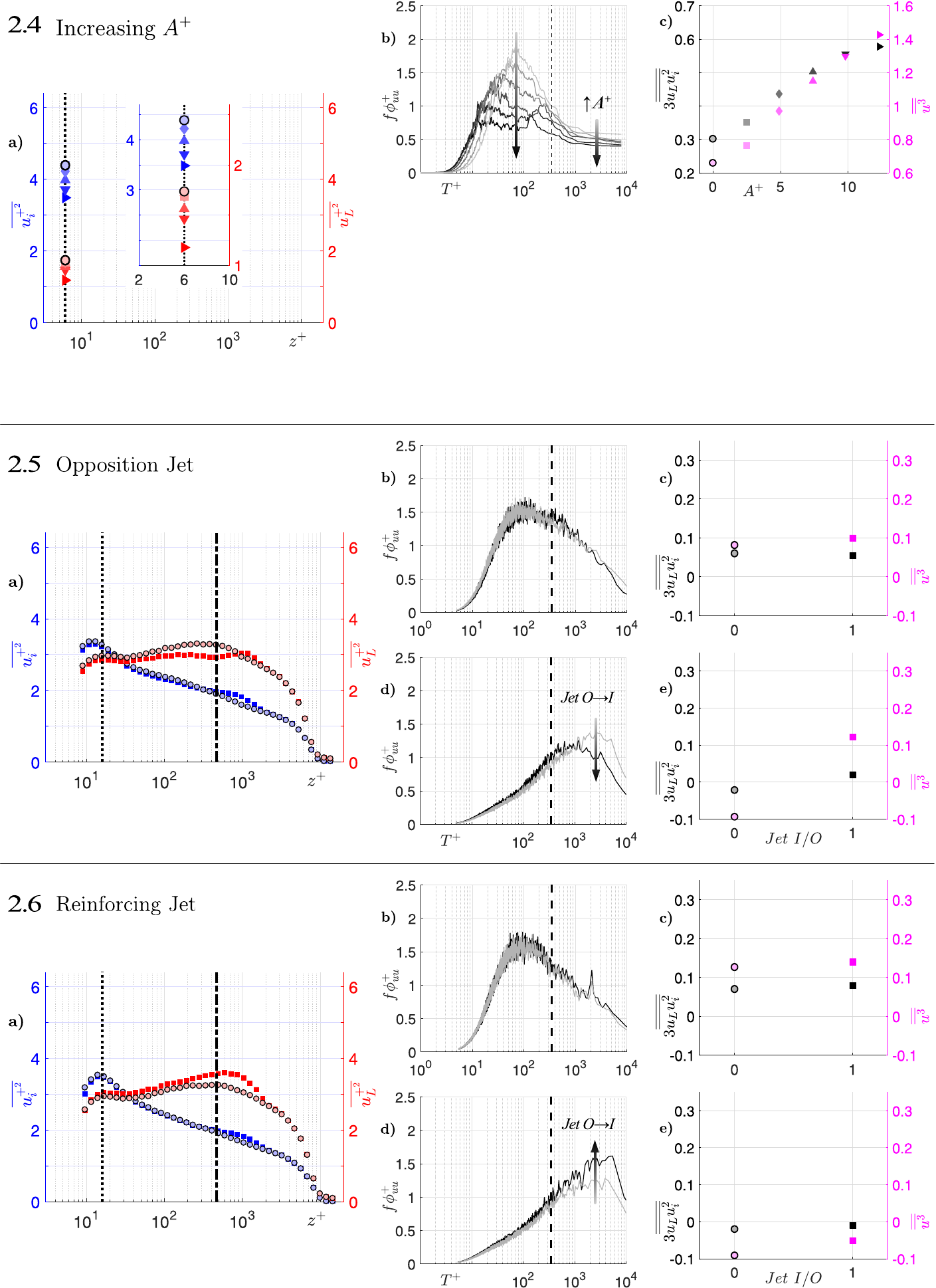}
\caption{\label{Fig2b} 
(\emph{continued}) $-$ (\emph{left column}) Variance of inner- (left axis) and large- (right axis) scale velocity fluctuations for varying intensities of the following perturbations: \textbf{2.4} Increasing $A^+$, \textbf{2.5} Opposition jet and \textbf{2.6} Reinforcing jet. 
(\emph{center column}) Premultiplied energy spectra for $z^+$-locations corresponding to (b) the inner region (indicated by a vertical dotted line in (a)) and (d) the outer region (indicated by a dash-dotted line in (a)). 
Vertical dashed lines in the center column represent the cut-off timescale, $T^{+}_{c}=350$. 
(\emph{right column}) Cross-term, $3\overline{\overline{u_{L} u_{i}^{2}}}$, (left axis) and skewness of streamwise velocity fluctuations, $\overline{\overline{u^{3}}}$ (right axis) as a function of increasing perturbation intensities at $z^+$-locations, matched with (b,d) respectively, corresponding to (c) the inner region and (e) the outer region. 
Reference canonical cases are shown in the lightest shading, and symbols for the reference canonical cases have a solid black outline.}
\end{figure}

\begin{figure}
\includegraphics[width=0.89\textwidth]{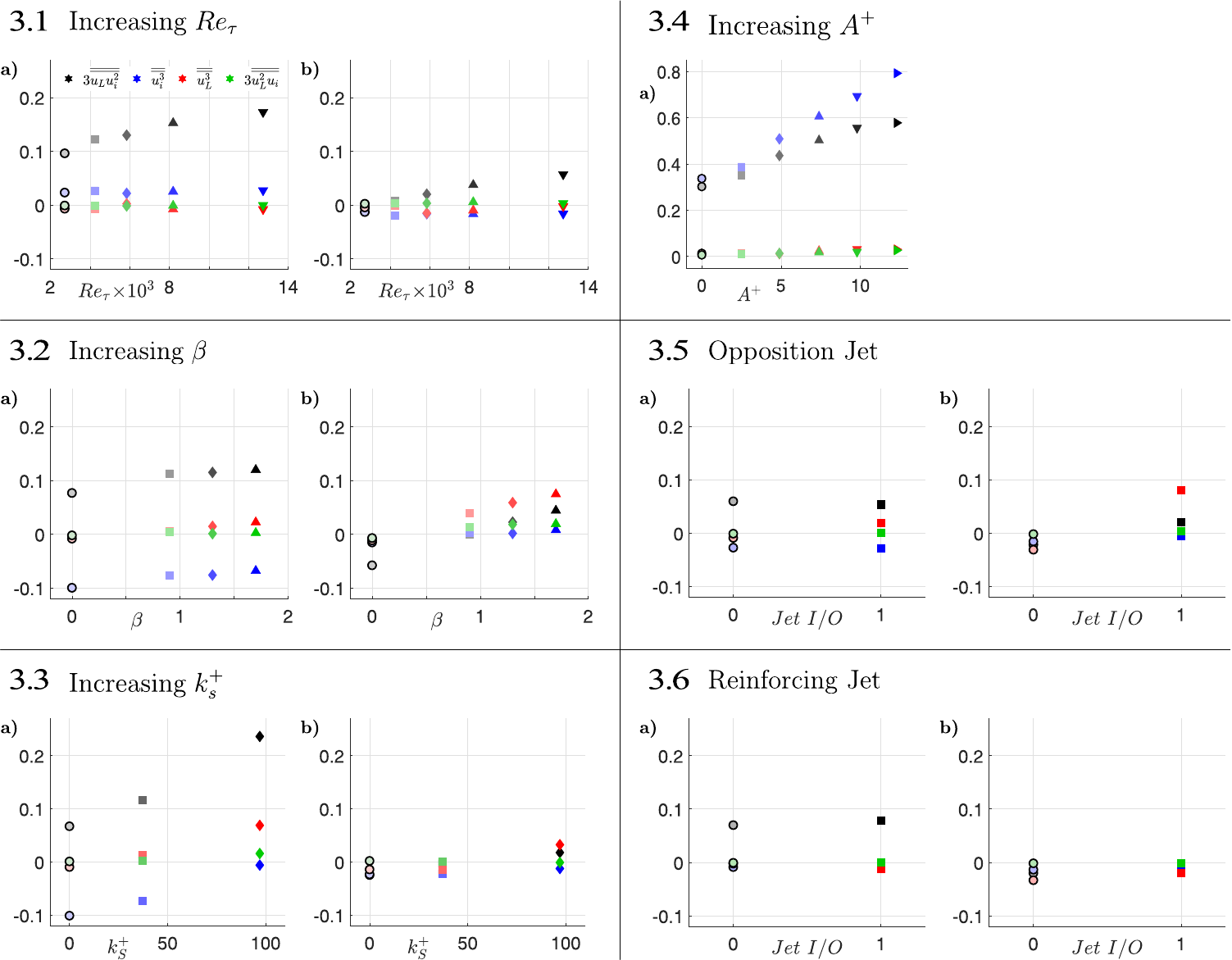}
\caption{\label{Fig3}
Cross-terms ($3\overline{\overline{u_{L} u_{i}^{2}}}$, $3\overline{\overline{u_{L}^{2} u_{i}}}$) and skewness of inner- and large-scale velocity fluctuations ($\overline{\overline{u_{i}^{3}}}$, $\overline{\overline{u_{L}^{3}}}$) as a function of increasing perturbation intensities at $z^+$-locations corresponding to (a) the inner region (vertical dotted lines) and (b) the outer region (dash-dotted lines) as indicated in the corresponding subfigures (a) in figure~\ref{Fig2a}. 
Results from reference canonical cases are shown in the lightest shading, and have a solid black outline.}
\end{figure}

\subsection{Increasing APG strength, $\beta$}

The second dataset is from Deshpande et al. \cite{deshpande_reynolds-number_2023} where the perturbation is a low-to-moderate adverse-pressure gradient (APG) with increasing strength as quantified by the Clauser pressure-gradient parameter, $\beta$. 
Here, ${\beta}(x)$ $=$ $(\delta^* / {\rho U_\tau^2}) \left( {\rm d} P/ {\rm d} x \right)$, where $\delta^* = \int_0^{\delta}(1 - U(z)/U_e) {\rm d}z$ is the displacement thickness, $U(z)$ is the mean streamwise velocity at wall-normal location $z$, $U_e$ $=$ $U$($z$ = $\delta$), \emph{i.e.} the edge velocity, $\rho$ is the fluid density and ${\rm d} P/{\rm d} x$ is the mean streamwise pressure-gradient at the measurement location $x$. 
In this experiment, increasing $\beta$ values were achieved by systematically increasing the number of low-porosity screens installed at the tunnel outlet. 
The presence of these screens increased the tunnel static/back pressure, while openings along the test section roof permitted the pressurized air to bleed out from various streamwise locations, thereby imposing a moderately strong APG on the TBL developing along the bottom wall \citep{deshpande_reynolds-number_2023}. 
A single hot-wire measurement station (located near the downstream end of the test section to ensure a nominal high-$\re$ condition) was used, and the free-stream velocity at this location was matched for various $\beta$-cases to ensure a nominally matched $\re$. 
Based on the conclusions of Deshpande et al. \cite{deshpande_reynolds-number_2023}, such a perturbation is expected to increase the energy of both the inner- and large-scale velocity fluctuations, especially in the outer-region of the TBL. 

Velocity statistics associated with the various APG cases are presented in figures~\ref{Fig2a}.2. 
In figure~\ref{Fig2a}.2a, a distinct increase in $\overline{u_{i}^{+^{2}}}$ can be seen in the outer region for all APG cases (relative to the ZPG case), while it does not change significantly in the inner region.
This is well-known in the literature \citep{pozuelo_adverse-pressure-gradient_2022,deshpande_reynolds-number_2023} for weak and moderately strong APG TBLs, with the behaviour in the outer region being associated with the migration of the near-wall `inner' scales to the outer region \citep{vinuesa_turbulent_2018}. 
Changes in $\overline{u_{L}^{+^{2}}}$ can also be seen across all $z^{+}$-locations, specifically an increase in the outer region ($z^{+}>200$) can be noted that is proportional to $\beta$ \citep{pozuelo_adverse-pressure-gradient_2022,deshpande_reynolds-number_2023}. 
The premultiplied spectra in figure~\ref{Fig2a}.2b are from $z^{+}=24$, as indicated by the vertical dotted line in figure~\ref{Fig2a}.2a. 
This is the $z^{+}$-location nearest to the wall where data was available for each case, including the baseline canonical case. 
The only discernible change is an increase in the premultiplied spectra amplitude for large scales (right of vertical dotted line at $T^{+}_{C}=350$), which is associated with an increase in $\beta$. 
The premultiplied spectra in figure~\ref{Fig2a}.2d are from $z^{+}=1000$, where large pressure gradient related changes to the variance are observed, indicated by the vertical dash-dotted line in figure~\ref{Fig2a}.2a. 
At this $z^{+}$-location there is a broadband increase in the premultiplied spectra amplitude, corresponding with an increase in $\beta$, however the amplitude of inner scales ($T^+$ $<$ $T^{+}_{C}$) does not change significantly with increasing $\beta$ for $\beta \gtrsim 0.9$, similar to the trend seen in $\overline{u_{i}^{+^{2}}}$ in figure~\ref{Fig2a}.2a.

The effect of increasing $\beta$ on the skewness and the cross-term has been documented in figures~\ref{Fig2a}.2c,e. 
Both these terms exhibit net growth with increasing $\beta$, at the inner as well as outer $z^{+}$-locations. 
This effect can be seen for all APG strengths in the dataset, ranging from a relatively weak APG ($\beta<1$, square symbols) to a moderate APG ($\beta\sim1.7$, triangle symbols). 
However, unlike the case of canonical TBLs, there are noticeable differences between the magnitudes of the skewness and the cross-term, and their rate of increase with $\beta$, suggesting changes in triadic interactions apart from those associated with the amplitude modulation phenomena. 
This is evidenced in figures~\ref{Fig3}.2a,b where both $\overline{\overline{u_{L}^{3}}}$ and $\overline{\overline{u_{i}^{3}}}$ are noted to increase with $\beta$, particularly rapidly in the outer region. 
This enhancement can be associated with the energization of both the inner and large scales by the APG, as discussed above \citep{pozuelo_adverse-pressure-gradient_2022,vinuesa_turbulent_2018}. 
It also confirms the influence of APG on the non-linear interactions spanning across the entire scale hierarchy coexisting in the TBL. 
These results suggest that, within the framework of increasing APG strength ($\beta$) as a perturbation to the baseline canonical TBL, enhancement of all triadic couplings need to be accounted for in any predictive models for APG TBLs. 
\rev{Moreover, the uni-directionality associated with the amplitude modulation phenomena (\emph{i.e.} $u_{L}$ governing inter-scale interactions) cannot be extended to APG TBLs given both $u_{L}$ and $u_{i}$ are manipulated with increasing $\beta$} (figure~\ref{Fig2a}.2a). \\

\subsection{Increasing surface roughness, $k^+_s$}

The third dataset is from Squire et al. \cite{squire_comparison_2016}, where the perturbation is increasing surface roughness characterized by an increasing Nikuradse-roughness height, $k_{s}^{+}$. 
In this experiment, the entire tunnel floor was covered with a P36 grit sandpaper, while both the free-stream velocities and streamwise measurement locations were adjusted such that a nominally constant high-$\re$ was maintained across various $k_{s}^{+}$ cases. 
This also meant that ${k_s}/{\delta}$ increased with $k^+_{s}$, however, the relative roughness height consistently remained small across all cases (\emph{i.e.} ${\delta}/{k_s}$ $\gtrsim$ 40). 

Based on the conclusions of Squire et al. \cite{squire_comparison_2016}, surface roughness is expected to decrease the energy of both the inner- and large-scale velocity fluctuations near the wall. 
As such, this perturbation directly affects the TBL region (I) labelled in figure~\ref{Fig0}, and this is confirmed from the hot-wire statistics presented in figures~\ref{Fig2a}.3. 
In figure~\ref{Fig2a}.3a, there is a significant decrease in $\overline{u_{i}^{+^{2}}}$ with increasing $k_{s}^{+}$ (light to dark colors) across all $z^{+}$-locations. 
There is also a decrease in $\overline{u_{L}^{+^{2}}}$, which is primarily localized to the near-wall region ($z^{+}<100$). 
The premultiplied spectra in figure~\ref{Fig2a}.3b are plotted for $z^{+} \approx 40$, which was the nearest possible location for hot-wire data acquisition among all roughness cases considered. 
The amplitude of the premultiplied spectra is noted to decrease across all scales (\emph{i.e.} all $T^{+}$) with an increase in $k_{s}^{+}$. 
This confirms that there is a decrease in the energy of both $u_{i}$ and $u_{L}$ near the wall, corresponding with the increase in $k_{s}^{+}$. 
The premultiplied spectra in figure~\ref{Fig2a}.3d are plotted for $z^{+}=400$ (\emph{i.e.} in the log region). 
At this location there is still a decrease in the energy of inner scales (left of vertical dotted line at $T^{+}_{C}=350$), but there is now an increase in the energy of large scales (right of vertical dotted line at $T^{+}_{C}=350$), associated with the largest $k_{s}^{+}$. 
These spectra highlight the unique aspects of the surface roughness perturbation which set it apart from the previous two cases. 

The effect of increasing $k_{s}^{+}$ on the skewness and the cross-terms are shown in figures~\ref{Fig2a}.3c,e. 
Both the skewness and the cross-term show a positive change with increasing $k_{s}^{+}$, which can be seen across both transitionally rough (0 $<$ $k_{s}^{+}$ $<$ 80; square symbols) and fully-rough scenarios ($k_{s}^{+}$ $>$ 80; diamond symbols), at both $z^{+}$-locations. 
These results suggest that, within the framework of increasing surface roughness ($k_{s}^{+}$) as a perturbation to the baseline canonical TBL, a decrease in energy of both inner and large scales is associated with a local increase in the magnitude of inter-scale coupling near the wall. 
Similarly, a decrease in energy of inner scales and an increase in the energy of large scales is also associated with a local increase in the magnitude of inter-scale coupling in the log region. 
Although this increase in the cross-term near the wall resembles an amplitude modulation effect, it differs fundamentally from the original definition by \citet{mathis_large-scale_2009} for a canonical TBL. 
In the case of increasing $k^+_s$, notably, magnitudes of both $u_{i}$ and $u_{L}$ are reduced in the near-wall region, and thus the increase in the cross-term cannot solely be attributed to the influence of the large scales. 
Further, increasing $k^+_s$ is found to influence the non-linear interactions spanning the entire energy spectrum, especially near the wall, which is evident from the various decomposed terms of skewness plotted in figure~\ref{Fig3}.3a,b. 
\rev{Therefore, akin to the conclusions regarding the effect of increasing $\beta$, the current analysis indicates the need to account for all non-linear triadic interactions in the development of predictive models for rough wall TBLs. 
Moreover, the growth of the cross-term cannot be characterized as an amplitude modulation effect in the traditional sense as defined by Mathis et al.} \citep{mathis_large-scale_2009}. 

Although the present analysis limits itself to scenarios of TBL developing over homogeneously distributed surface roughness, further interesting flow characteristics can be expected if step changes in surface roughness are considered. 
Interested readers are referred to the recent study by Li et al. \citep{li2023quantifying}, who have experimentally quantified the cross-term (\emph{i.e.}, amplitude modulation term) for a TBL developing after a rough-to-smooth change in wall condition. 
Their investigation revealed higher magnitudes of the cross-term in the near-wall region (relative to a canonical TBL), which was attributed to the more energetic footprints of the large-scale motions influenced by the upstream rough-wall condition. \\

\subsection{Increasing spanwise wall forcing, $A^+$}

The fourth dataset considered here is from Deshpande et al. \cite{deshpande_relationship_2023} where the perturbation is a spanwise oscillating wall with increasing magnitude of the spanwise wall velocity, given by $A^{+}=A/U_{\tau}$. 
In this experiment, forty-eight individual slats mounted along a 2.4\:m portion of the test section floor were oscillated in a phase-synchronized manner along the spanwise direction, leading to the creation of an $8\lambda$ long upstream-traveling sinusoidal wave with user specified frequencies (${f_{osc}}$). 
\rev{With a nominal stroke length ($d=18\:mm$) for the slat movement, the amplitude of the spanwise wall velocity ($A=2\pi{d}{f_{osc}}$) is directly related to the user varied oscillation frequency.} 
For the experiments considered in this study, the wall oscillation frequency range was maintained between 0.000 ${\le}\;{{f}^{+}_{osc}}\;{\le}$ 0.007, which led to a variation in spanwise wall forcing across 0.0 ${\le}\;{{A}^{+}}\;{\le}$ 12.3. 
Hot-wire measurements were conducted in the very near-wall region, above the oscillating slats, for both the non-actuated and actuated cases. 
The wind tunnel free-stream velocity was matched in all cases. 

Based on the findings of Deshpande et al. \cite{deshpande_relationship_2023}, this wall actuation attenuates the energy of both inner- and large-scale velocity fluctuations, thereby affecting the energy spectra across regions (I)$-$(III) of figure~\ref{Fig0}. 
\rev{The hot-wire statistics from these cases are presented in figures~\ref{Fig2b}.4, which although limited to a single $z^+$-location, exhibit trends consistent with other locations in the near-wall region (confirmed by comparisons with independent PIV measurements in \citep{deshpande_near-wall_2023}, which show wall-normal profiles across most of the TBL).} 
A seen in figure~\ref{Fig2b}.4a, a significant decrease in both $\overline{u_{i}^{+^{2}}}$ and $\overline{u_{L}^{+^{2}}}$ with increasing $A^+$ (light to dark colors) at $z^{+}=6$ is observed. 
This is confirmed by figure~\ref{Fig2b}.4b, which shows an attenuation of the amplitude of the premultiplied spectra across all $T^{+}$ $\gtrsim$ 50, with increasing $A^{+}$. 
The effect of increasing $A^{+}$ on the skewness and the cross-term at $z^{+}=6$ is shown in figure~\ref{Fig2b}.4c, where both these statistics are found to exhibit a significant positive change with increasing $A^{+}$ (\emph{i.e.}, increasing drag reduction \citep{deshpande_relationship_2023,deshpande_near-wall_2023}). 
Similar observations of increasing skewness, with increasing drag reduction, have been noted previously on introduction of local oscillating blowing \citep{tardu2001active} or microbubbles and polymers \citep{pal1989comparison} in the TBL. 
However, none of these studies associated their observations with inter-scale interactions, which has been discussed here. 
A notable observation in the present study is the gradual positive change in the non-linear interactions, starting from the smallest perturbation/$A^{+}$ (in square symbols), through to the largest perturbation/$A^{+}$ (right pointing triangle symbols). 
These results suggest that, within the framework of increasing spanwise wall forcing as a perturbation to the baseline canonical TBL, the decrease in near-wall energy of both inner and large scales is accompanied with a significant local increase in the inter-scale coupling. 
\rev{However, similar to the conclusions drawn about the previous non-canonical perturbations, the present increase in the cross-term cannot be associated with the amplitude modulation defined in the classical sense, given it is a consequence of attenuation of both large and inner scales. 
Further, significant differences between the magnitudes, and rate of increase, of the cross-term and skewness in figure~\ref{Fig2b}.4c can be explained by the increasing magnitude of the $\overline{\overline{u_{i}^{3}}}$ term (with increasing $A^+$) as shown in figure~\ref{Fig3}.4a, but not of $\overline{\overline{u_{L}^{3}}}$. 
This reaffirms the recurring conclusion that all triadic interactions should be considered when modeling the effect of forcings, such as spanwise wall oscillation, on the TBL. 

Collectively, the aforementioned cases reveal the uniqueness with which different non-canonical perturbations can affect the hierarchy of triadic interactions coexisting in the TBL. 
Additionally, these perturbations all typically resulted in a decrease in phase between coupled scales on average (\emph{i.e.} increasing skewness), but this increase was shown to not be a function of amplitude modulation alone (\emph{i.e.} additional terms of Eq~\ref{eq:skew2} become significant). 
These results also motivate the consideration of all triadic interactions when modeling wall-bounded flows, particularly in the case of non-canonical flows.} \\

\subsection{Opposing and reinforcing wall-normal jet forcing}

The fifth and sixth datasets are from Abbassi et al. \cite{abbassi_skin-friction_2017}, where the perturbation is a wall-normal jet with a control scheme designed to either enhance (reinforcing case) or weaken (opposition case) large-scale velocity fluctuations in the outer region. 
In these experiments, a spanwise array of hot-film sensors placed on the wall, upstream of the jet-actuators, were used to identify the footprint of incoming $u_{L}$ motions as an input for the actuation scheme. 
The actuators were composed of a spanwise array of streamwise elongated slots in the wall, which generated a wall-normal jet of pressurized air supply that penetrated into the log region of the TBL. 
A single hot-wire measurement station downstream of the jets was used, and the free-stream velocity was matched in both the actuated and non-actuated cases to maintain a nominally similar high-$\re$ inflow condition. 
In the opposition control case, the hot-film sensors were used to identify incoming high-energy ($+u_{L}$) large-scale motions so the jets could fire against these high-skin-friction-contributing regions \citep{abbassi_skin-friction_2017}. 
On the other hand, the same system was used to identify and target incoming low-energy ($-u_{L}$) large-scale motions for the reinforcing case. 

Abbassi et al. \citep{abbassi_skin-friction_2017} found the opposing and reinforcing cases to respectively attenuate and enhance the large-scale energy in the outer region. 
This is indeed depicted by $\overline{u_{L}^{+^{2}}}$ presented in figures~\ref{Fig2b}.5a and \ref{Fig2b}.6a for the opposition and reinforcing cases respectively, while $\overline{u_{i}^{+^{2}}}$ remains relatively unaffected for both. 
Hence, both these cases represent a unique scenario where a change in the large-scale energy in the outer region is not reflected in their signatures/footprints close to the wall. 
This is supported by the premultiplied spectra plotted for both the reinforcing (figure~\ref{Fig2b}.5b) and opposing cases (figure~\ref{Fig2b}.6b) compared against their corresponding non-actuated cases at $z^{+} \approx 15$. 
Consequently, the effects of either of these wall-normal jet forcings, on the skewness and the cross-terms, are negligible at $z^{+}=15$ (depicted in figures~\ref{Fig2b}.5c and \ref{Fig2b}.6c). 
Alternatively, the premultiplied spectra plotted for the reinforcing (figure~\ref{Fig2b}.5d) and opposing cases (figure~\ref{Fig2b}.6d) at $z^{+} \approx 500$ (\emph{i.e.} in the log region) show a decrease or increase in the energy of large scales (right of vertical dotted line at $T^{+}_{C}=350$) corresponding with the forcing scheme and reflected in $\overline{u_{L}^{+^{2}}}$ in figures~\ref{Fig2b}.5a,6a. 
Now at $z^{+} \approx 500$ the effects of these wall-normal jet forcings, especially the opposition jet, on the skewness and the cross-terms, are discernible in figures~\ref{Fig2b}.5e,6e). 
Further, figures~\ref{Fig3}.5 and \ref{Fig3}.6 demonstrate that the $\overline{\overline{u_{L}^{3}}}$ term of Eq. (\ref{eq:skew2}) is most strongly influenced by the jet forcings, while there is a negligible influence on any other non-linear couplings, especially in the near-wall region (in a statistically-averaged sense). 
Hence, both the present jet forcing cases suggest the possibility of a unique scenario where near-wall non-linear couplings remain unaffected despite the large-scale fluctuations in the outer region being energized/attenuated. 
\rev{It suggests that not every large-scale phenomena in the outer region influences the near-wall region, and hence the prediction of the near-wall signatures based on the outer motions may not be as straightforward for such complex non-canonical effects (as was for canonical TBLs).} \\

\section{Summary and conclusions}
\rev{This study evaluates the connection between the amplitude modulation phenomenon, observed by Mathis et al. \citep{mathis_large-scale_2009} for canonical wall-bounded flows, and the inter-scale interactions observed in non-canonical TBLs.} 
Published hot-wire datasets corresponding to various non-canonical effects are considered, including surface roughness, adverse-pressure gradients, spanwise and wall-normal forcing, each of which are associated with unique changes in the inner- and large-scale energy (relative to their canonical baseline cases). 
These changes, as documented in table \ref{Tab2}, are responsible for distinguishing their inter-scale interactions from the classical definition of amplitude modulation conceived by Mathis et al. \citep{mathis_large-scale_2009}. 
\rev{As such, interpretation of the correlations conventionally associated with amplitude modulation (\emph{i.e.} $3\overline{\overline{u^{2}_{i}u_{L}}}$ and/or $R$) should be considered more cautiously in the future, particularly in the context of non-canonical wall-bounded flows. 
The present study also demonstrates a clear distinction between both the range of triadically-coupled scales, and wall-normal regions of the TBL, which are influenced by different non-canonical perturbations (table \ref{Tab2}). 
These results underscore the necessity of quantifying interactions across a broad scale hierarchy, via dedicated two-point correlations for example, when constructing predictive models for non-canonical wall-bounded flows.}

\begin{table}
\caption{\label{Tab2} Summary of key findings based on figures \ref{Fig2a} and \ref{Fig3}. $\uparrow$ and $\downarrow$ respectively indicate increase or decrease in magnitude relative to their canonical baseline cases.}
\begin{ruledtabular}
\begin{tabular}{cccccccc}
\vspace{2mm}
Perturbation & TBL region & $\overline{u^2_i}$ & $\overline{u^2_L}$ & $\overline{\overline{{u^{2}_{i}}{u_{L}}}}$ 
 & $\overline{\overline{u_{i}^{3}}}$ & $\overline{\overline{u_{L}^{3}}}$ & $\overline{\overline{u_{L}^{2}u_{i}}}$\\
\hline
\multirow{2}{*}{Increasing $Re_{\tau}$} &
      inner & $\approx$ & $\uparrow$ & $\uparrow$ & $\approx$ & $\approx$ & $\approx$\\
      & outer & $\approx$ & $\uparrow$ & $\uparrow$ & $\approx$ & $\approx$ & $\approx$\\
\hline      
\multirow{2}{*}{Increasing $\beta$} &
      inner & $\approx$ & $\uparrow$ & $\uparrow$ & $\uparrow$ & $\uparrow$ & $\approx$\\
      & outer & $\uparrow$ & $\uparrow$ & $\uparrow$ & $\uparrow$ & $\uparrow$ & $\approx$\\  
\hline      
\multirow{2}{*}{Increasing $k^{+}_{s}$} &
      inner & $\downarrow$ & $\downarrow$ & $\uparrow$ & $\uparrow$ & $\uparrow$ & $\approx$\\
      & outer & $\downarrow$ & $\uparrow$ & $\uparrow$ & $\approx$ & $\uparrow$ & $\approx$\\
\hline
Increasing $A^{+}$ & inner & $\downarrow$ & $\downarrow$ & $\uparrow$ & $\uparrow$ & $\approx$ & $\approx$\\
\hline
\multirow{2}{*}{Opposing $u_{L}$} &
      inner & $\approx$ & $\approx$ & $\approx$ & $\approx$ & $\approx$ & $\approx$\\
      & outer & $\approx$ & $\downarrow$ & $\uparrow$ & $\approx$ & $\uparrow$ & $\approx$\\
\hline
\multirow{2}{*}{Reinforcing $u_{L}$} &
      inner & $\approx$ & $\approx$ & $\approx$ & $\approx$ & $\approx$ & $\approx$\\
      & outer & $\approx$ & $\uparrow$ & $\uparrow$ & $\approx$ & $\uparrow$ & $\approx$\\
\end{tabular}
\end{ruledtabular}
\end{table}

Interestingly, it was observed that several (though not all) perturbation effects led to an enhancement of inter-scale coupling and broadband changes in the velocity spectra, with increasing perturbation intensities (table \ref{Tab2}). 
This observation is consistent with past studies \citep{duvvuri2016nonlinear,deshpande_relationship_2023,deshpande_near-wall_2023} and highlights the tendency of turbulence scales to align more closely (\emph{i.e.}, exhibit reduced phase lag) when subjected to certain external forcings/perturbations. 
In terms of flow physics, the manipulation of inter-scale coupling in non-canonical flows could be feasibly linked with changes in scale-dependent inclination angles of coherent structures \citep{deshpande2019streamwise,cui2024variability}. 
Previous research in canonical flows \citep{jacobi_phase_2013,cui2024variability} has already indicated these angles to be governing inter-scale phase relationships. 
Alternatively, unique perspectives such as the QSQH theory \citep{zhang2016qsqh, chernyshenko2021extension} may also be able to explain some of the changes in inter-scale interactions noted here for non-canonical TBLs. 
This collectively encourages a more thorough exploration of wall turbulence from a dynamical systems perspective \citep{duvvuri_triadic_2015,duvvuri2016nonlinear}.

\bigskip
\noindent We gratefully acknowledge A/Prof. Ian Jacobi for insightful discussions, and the assistance of Dr. Reza Abbassi and Dr. Dougal Squire with regards to accessing the published experimental datasets. 
R. D. is grateful for financial support from the University of Melbourne's Postdoctoral Fellowship. 
R. D. and I. M. gratefully acknowledge funding from the Office of Naval Research (ONR) and ONR global grant: N62909-23-1-2068. \\

\section*{Author contributions}
\noindent \textbf{Mitchell Lozier:} Writing $-$ original draft, Investigation, Methodology, Validation, Formal analysis.\\ 
\textbf{Ivan Marusic:} Conceptualization, Writing $-$ review \& editing, Funding acquisition, Supervision.\\ 
\textbf{Rahul Deshpande:} Conceptualization, Methodology, Writing $-$ review \& editing, Funding acquisition, Supervision.

\bibliography{references}

\end{document}